\documentclass[journal]{IEEEtran}

\usepackage[cmex10]{amsmath}

\usepackage[caption=false,font=footnotesize]{subfig}

\usepackage{cite,color,amsfonts}
\usepackage{algorithm}
\usepackage{algpseudocode}
\usepackage{enumerate}

\ifCLASSINFOpdf
  \usepackage[pdftex]{graphicx}
  \graphicspath{{./fig/}}%
  \DeclareGraphicsExtensions{.png,.pdf}
\else
  \usepackage[dvips]{graphicx}
  \graphicspath{{./eps/}}
  \DeclareGraphicsExtensions{.eps}
\fi

\newcommand{\bi}{\begin{itemize}}	
\newcommand{\ei}{\end{itemize}}	
\newcommand{\bn}{\begin{enumerate}}	
\newcommand{\en}{\end{enumerate}}
\newcommand{\bc}{\begin{center}}
\newcommand{\ec}{\end{center}}
\newcommand{\be}{\begin{equation}}
\newcommand{\ee}{\end{equation}}
\newcommand{\ben}{\begin{equation*}}
\newcommand{\een}{\end{equation*}}
\newcommand{\beqa}{\begin{eqnarray}}
\newcommand{\eeqa}{\end{eqnarray}}

\newcommand{\red}{\textcolor{black}}

\begin{document}%
\title{Belief consensus algorithms for fast distributed target tracking in wireless sensor networks}

\author{Vladimir~Savic, Henk Wymeersch, and~Santiago~Zazo
\thanks{V.\,Savic was with the Signal Processing Applications Group, Universidad Politecnica de Madrid, Madrid, Spain, and he is now with the Dept. of Electrical Engineering (ISY), Link\"{o}ping University, Sweden (e-mail: vladimir.savic@liu.se). H.\,Wymeersch is with the Dept. of Signals and Systems, Chalmers University of Technology, Gothenburg, Sweden (e-mail: henkw@chalmers.se). S.\,Zazo is with the Signal Processing Applications Group, Universidad Politecnica de Madrid, Madrid, Spain (e-mail: santiago@gaps.ssr.upm.es)}%
\thanks{This work is supported by the Swedish Foundation for Strategic Research (SSF) and ELLIIT; Swedish Research Council (VR), under grant no. 2010-5889; the European Research Council, under grant COOPNET No. 258418; program CONSOLIDER-INGENIO 2010 under the grant CSD2008-00010 COMONSENS; the European Commission under the grant FP7-ICT-2009-4-248894-WHERE-2; and the FPU fellowship from Spanish Ministry of Science and Innovation.  Part of this work was presented at the 2012 European Signal Processing Conference.}%
}

\maketitle 

\begin{abstract}
In distributed target tracking for wireless sensor networks, agreement on the target state can be achieved by the construction and maintenance of a communication path, in order to exchange information regarding local likelihood functions. Such an approach lacks robustness to failures and is not easily applicable to ad-hoc networks. To address this, several methods have been proposed that allow agreement on the global likelihood through fully distributed belief consensus (BC) algorithms, operating on local likelihoods in distributed particle filtering (DPF). However, a unified comparison of the convergence speed and communication cost has not been performed. In this paper, we provide such a comparison and propose a novel BC algorithm based on belief propagation (BP). 
According to our study, DPF based on metropolis belief consensus (MBC) is the fastest in loopy graphs, while DPF based on BP consensus is the fastest in tree graphs. Moreover, we found that BC-based DPF methods have lower communication overhead than data flooding when the network is sufficiently sparse.
 
\end{abstract}



\section{Introduction}\label{sec:intro}
Distributed tracking in wireless sensor networks (WSN) \cite{Hlinka2013} is an important task for many applications in which central unit is not available. For example, in emergency situations, such as a fire, a nuclear disaster, or a mine collapse, a WSN can be used to detect these phenomena. Once a phenomenon is detected (e.g., increased temperature or radioactivity), the sensors start to sense their neighbourhood and cooperatively track people and assets. As sensors are low-cost devices that may not survive during tracking, it is important to track in a manner that is fully robust to sensors failures, and in such a way that every sensor has the same belief of the target location. Then, the rescue team can access the estimates, even if just one sensor survives. As another potential application, sensor nodes can also serve as actuators, which perform a specific action (e.g., move towards the target) as a function of estimated target's position. In this case, to ensure compatible actions, a unified view of the target's position is crucial.

The traditional approach to target tracking is based on Kalman filtering (KF) \cite{Welch2006}. However, due to nonlinear relationships and possible non-Gaussian uncertainties,  a particle filter (PF) is preferred \cite{Arulampalam2002} in many scenarios. Therefore, the focus of this paper will be on PF-based distributed tracking. Many PF-based methods are based on the construction and maintenance of a communication path, such as a spanning tree or a Hamiltonian cycle. For example, in \cite{Coates2004}, low-power sensors pass the parameters of likelihood functions to high-power sensors, which are responsible to manage the low-power nodes. In \cite{Sheng2005}, a set of uncorrelated sensor cliques is constructed, in which slave nodes have to transmit Gaussian mixture parameters to the master node of the clique. The master node performs the tracking, and forward estimates to another clique. In \cite{Lee2009}, a Markov-chain distributed PF  is proposed, which does not route the information through the graph during tracking. However, it requires that each node knows the total number of communication links and the number of communication links between each pair of nodes, which can be obtained only by aggregating the data before tracking. In \cite{Hlinka2009}, the authors propose an incremental approach, in which the parameters of the likelihood are communicated from sensor to sensor in order to approximate the posterior of interest. Finally, there is also a different class of methods \cite{Coates2005,Jiang2011} that maintain disjoint sets of particles at different nodes, and propagate them towards the predicted target position. These type of methods, also know as \textit{leader-agent algorithms} (see \cite{Hlinka2013} for an overview), lack robustness to failures, cause excessive delays due to the sequential estimation, and do not provide the estimates at each sensor without additional post-processing routing phase.

These problems can be solved if each node broadcasts observations until all the nodes have complete set of observations. Then, each node (acting like a fusion center) performs the tracking. This method, known as \textit{data flooding} and used in non-centralized PF (NCPF) \cite{Djuric2011a}, is not scalable, but can be competitive in some scenarios. Other solutions consider distributed particle filtering (DPF) methods based on consensus algorithms \cite{Gu2007a,Gu2008,Hlinka2011,Hlinka2012,Oreshkin2010,Ustebay2011,Farahmand2011}. In \cite{Gu2007a}, the global posterior distribution is approximated with a Gaussian mixture, and consensus is applied over the local parameters to compute the global parameters. Similarly,  \cite{Gu2008,Hlinka2011} use a Gaussian approximation instead of a Gaussian mixture, and \cite{Hlinka2012} can use any distribution that belongs to an exponential family. Randomized gossip consensus was used in \cite{Oreshkin2010} for distributed target tracking.  The main problem with these approaches is that the global likelihood function is represented in the same parametric form as local likelihood functions, which is questionable in certain scenarios. In \cite{Ustebay2011, Farahmand2011}, consensus is applied instead to the weights in the DPF, so that any likelihood can be represented. However, an issue that arises with these DPF approaches is that  consensus can be slow. In a setting where the target moves, only a finite time is available to perform consensus \cite{Lindberg2013}, so the fastest possible method should be employed. A recent and detailed overview of DPF algorithms can be found in  \cite{Hlinka2013}, but it does not analyze the effect of different consensus techniques on convergence.

In this paper, we compare five algorithms for target tracking using distributed particle filtering (DPF) based on belief consensus (BC):
\begin{enumerate}
\item standard belief consensus (SBC) \cite{Farahmand2011};
\item randomized gossip (RG) \cite{Ustebay2011};
\item broadcast gossip (BG) \cite{Liu2009}; 
\item Metropolis belief consensus (MBC) \cite{Hlinka2012}; and 
\item one novel algorithm based on belief propagation (BP), which we earlier proposed in \cite{Savic2012eusip}.
\end{enumerate}

To the best of our knowledge, this is the first study where these methods are compared in a common setting. According to our simulation study, DPF-MBC is the fastest in loopy graphs, while DPF-BP is the fastest in tree graphs (typical for tunnel-like environments). Moreover, we found that BC-based DPF methods have lower communication overhead than data flooding only in sparse networks.

The rest of this paper is organized as follows. In Section \ref{sec:overview}, we review centralized target tracking. In Section \ref{sec:dtrack}, we describe five BC algorithms for PF-based distributed target tracking, including the novel based on BP. Simulation results are shown in Section \ref{sec:sim}. Finally, Section \ref{sec:conc} provides our conclusions and suggestions for future work.

\section{Overview of Centralized Target Tracking}\label{sec:overview}
We assume that there is a number of static sensor nodes with known positions and one moving target (e.g., a person or vehicle) in some surveillance area. The target may be passive or not willing to reveal its state, but the sensors are assumed to periodically make observations that depend on their relative position to the target. The goal of the WSN is to track the state (e.g., position and velocity) of the target. In this section, we describe a centralized approach to solve this problem, in which all the observations are collected by a sensor that acts as a  fusion center. 

\subsection{System Model}\label{subsec:smodel}

\begin{figure}[!t]
\centering
\includegraphics[width=0.48\textwidth]{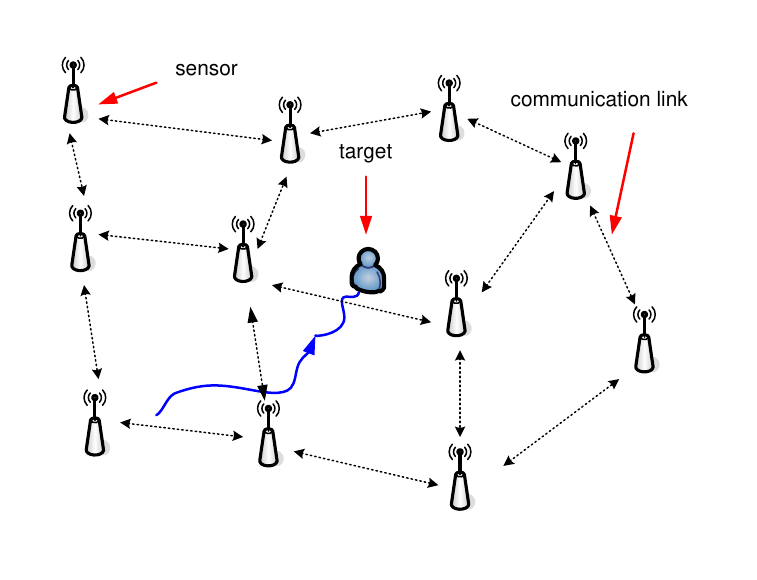} 
\caption{Illustration of target tracking in a WSN. The goal of the WSN is to track the position and velocity of the target.}
\label{fig:tracking-wsn}
\end{figure}

The scenario under consideration is illustrated in Figure~\ref{fig:tracking-wsn}. There are $N_s$ static sensors with known two-dimensional (2D) positions, $\mathbf{l}_n$ ($n=1, 2, \ldots, N_s$) and one mobile target with an unknown state $\mathbf{x}_t$ at time \emph{t}. 
The goal of the WSN is to estimate $\mathbf{x}_t$ at each (discrete) time $t$.  We use the following state-space model: 
\be
\label{eq:motion1}
{\mathbf{x}_{t + 1}} = f(\mathbf{x}_t,\mathbf{u}_t)
\ee
\be
\label{eq:measure1}
{y_{n,t}} = g_n(\mathbf{x}_t,v_{n,t}),
\ee
where ${\mathbf{u}_t}$ is  process noise, ${y_{n,t}}$ is local observation of sensor $n$ at time $t$, and ${v_{n,t}}$ is its observation noise. We denote the aggregation of all observations at time $t$ by $\mathbf{y}_t$. The process noise $\mathbf{u}_t$ can be non-Gaussian, but since it is usually hard to measure \cite{Welch2006, Savic2011a}, we can assume a Gaussian approximation with sufficiently large variance, which is a common choice.

We denote by  $G_t$  the set of the nodes that have an observation available at time $t$, and by $G_n$ the set of all neighbors of node $n$ (irrespective of whether or not they have an observation).
The observation noise $v_{n,t}$ is distributed according to $p_v(\cdot)$, which is not necessarily Gaussian, and typically depends on the particular measurement technique (e.g., acoustic \cite{Ahmed2010}, RSS \cite{Oka2010}, RF tomography \cite{Chen2011}) and the environment. Observations at any time $t$ are assumed to be conditionally independent.

We assume that the network is connected (i.e., there is at least a single path between any two nodes), and undirected (if  node $n$ can receive a packet from node $u$, then node $u$ can receive a packet from node $n$). For simplicity, we assume ideal probability of detection for both sensing and communication range, but more complex models can easily be incorporated \cite{Ihler2005a}. In other words, a sensor can detect the target if the distance between them is less than a certain value $r$, and two sensors can communicate with each other if the distance between them is less than $R$. Taking into account that radio of a node is usually  more powerful than its sensing devices \cite{Jiang2011,Galstyan2004}, we assume $R\ge r$.

\subsection{Centralized Particle Filtering}\label{subsec:cpf}
We apply the Bayesian approach for this tracking problem and recursively determine the posterior distribution $p(\mathbf{x}_t|\mathbf{y}_{1:t})$ given the prior $p(\mathbf{x}_{t-1}|\mathbf{y}_{1:t-1})$, dynamic model $p(\mathbf{x}_t|\mathbf{x}_{t-1})$ defined by (\ref{eq:motion1}), and the likelihood function $p(\mathbf{y}_t|\mathbf{x}_t)$ defined by (\ref{eq:measure1}). We assume that $p(\mathbf{x}_0|\mathbf{y}_0)=p(\mathbf{x}_0)$ is initially available. The posterior can be found using the standard prediction and correction equations \cite{Arulampalam2002}:
\be
\label{eq:prediction}
p(\mathbf{x}_t |\mathbf{y}_{1:t - 1} ) = \int {p(\mathbf{x}_t |\mathbf{x}_{t - 1})p(\mathbf{x}_{t - 1} |\mathbf{y}_{1:t - 1} )\mathrm{d} \mathbf{x}_{t - 1}}
\ee
\be
\label{eq:filtering}
p(\mathbf{x}_t |\mathbf{y}_{1:t} ) \propto p(\mathbf{y}_t |\mathbf{x}_t )p(\mathbf{x}_t |\mathbf{y}_{1:t - 1}).
\ee 
Due to the conditional independence among observations at time $t$, the global likelihood function $p(\mathbf{y}_t |\mathbf{x}_t )$ can be written as the product of the local likelihoods:
\be
\label{eq:g-lhood}
p(\mathbf{y}_t |\mathbf{x}_t ) \propto \prod\limits_{n \in G_t} p(y_{n,t}|\mathbf{x}_t ),
\ee
where local likelihood  $p(y_{n,t}|\mathbf{x}_t )$ is a function of state $\mathbf{x}_t$ for a given observation $y_{n,t}$. For notational convenience we will still write $p(y_{n,t}|\mathbf{x}_t )$ for $n \notin G_t$, with the tacit assumption that this function is equal to 1. 
 
Since the observation noise is generally not Gaussian, and the observation is not a linear function of the state, a traditional Kalman filtering \cite{Welch2006} approach cannot be used. Instead, we apply the particle filter \cite{Arulampalam2002}, in which the posterior distribution is represented by a set of samples (particles) with associated weights. A well-known solution is the sample-importance-resampling (SIR) method, in which $N_p$ particles are drawn from $p(\mathbf{x}_t|\mathbf{x}_{t-1})$, then weighted by the likelihood function, $p(\mathbf{y}_t|\mathbf{x}_t)$, and finally resampled in order to avoid degeneracy problems (i.e., the situation in which all but one particle have negligible weights). More advanced versions of PF also exist \cite{Pitt1999,Merwe2001,Kotecha2003}, but we focus on SIR since the distributed implementation of most PF-based methods is similar. We will refer to PF with SIR  as centralized PF (CPF). The CPF method is summarized in Alg.~\ref{SIR}. The CPF must be initialized with a set of particles $\{w_{0}^{(m)},\mathbf{x}_0^{(m)}\}$ drawn from $p(\mathbf{x}_0)$.
\begin{algorithm} [!t]
\caption{Centralized PF (CPF) (at time \emph{t})}\label{SIR}
\begin{algorithmic}[1]
\ForAll {particles $m=1:N_p$}
\State Draw particle: $\mathbf{x}_t^{(m)} \sim p(\mathbf{x}_t |\mathbf{x}_{t - 1}^{(m)})$
\State Compute weight: $w_t^{(m)}=w_{t-1}^{(m)} \cdot p(\mathbf{y}_t |\mathbf{x}_t^{(m)})$
\EndFor
\State Normalize: $w_t^{(m)}=w_t^{(m)}/\sum\limits_{m'}{w_t^{(m')}}$ (for $m=1:N_p$)
\State Compute estimates: $\hat{\mathbf{x}}_{t}=\sum\limits_{m}w_{t}^{(m)}\mathbf{x}_{t}^{(m)}$
\State Resample with replacement from $\{w_{t}^{(m)},\mathbf{x}_{t}^{(m)}\}^{N_p}_{m=1}$
\end{algorithmic}
\end{algorithm}

This algorithm is run on one of the nodes in the WSN, which serves as fusion center. The main drawbacks of the CPF are \cite{Hlinka2012,Patwari2005}: i) large energy consumption of the nodes which are in proximity of the fusion center, ii) high communication cost in large-scale networks; iii) the posterior distribution cannot be accessed from any node in the network; and iv) the fusion center has to know the locations, observations, and observation models of all the nodes. In the following section we will focus on distributed implementations of PF method, which alleviate these problems.
\section{Distributed Target Tracking}\label{sec:dtrack}

Our goal is to track the target in a distributed way, such that all the nodes have a common view of the state of the target. We will first describe a flooding approach (the NCPF), followed by a general description of the DPF, which relies on a belief consensus algorithm that is run as an inner loop. We then describe 5 belief consensus methods, and finally quantify the communication cost of DPF and NCPF. 

\subsection{Non-Centralized Particle Filter}\label{subsec:ncpf}
In NCPF \cite{Djuric2011a}, every sensor broadcasts its observation, observation model and an identifier, along with observations, observation models and identifiers from neighbors. This flooding procedure is repeated until every node has access to all information, when a common posterior $p(\mathbf{x}_t|\mathbf{y}_{1:t})$ can be determined. 
This approach is not scalable, but can be competitive in some scenarios (see also Section \ref{subsec:comm-cost}).

\begin{algorithm} [!t]
\caption{Distributed PF (DPF) (at node \emph{n}, at time \emph{t})}\label{dist-SIR}
\begin{algorithmic}[1]
\ForAll {particles $m=1:N_p$}
\State Draw particle: $\mathbf{x}_{t}^{(m)} \sim p(\mathbf{x}_{n,t} |\mathbf{x}_{t - 1}^{(m)})$
\State Compute weight: $w_{n,t}^{(m)}=$
\Statex ~~~~$w_{t-1}^{(m)} \cdot \mathrm{BC}\left(p(y_{1,t} |\mathbf{x}_t^{(m)}),...,p(y_{N_s,t} |\mathbf{x}_t^{(m)})\right)$
\EndFor
\State Normalize: $w_{n,t}^{(m)}=w_{n,t}^{(m)}/\sum\limits_{m'}{w_{n,t}^{(m')}}$ (for $m=1:N_p$)
\State $\hat{w}_{t}^{(m)}= \mathrm{MC} \left(w_{1,t}^{(m)},...,w_{N_s,t}^{(m)} \right)$ (for $m=1:N_p$)
\State Normalize: $\hat{w}_{t}^{(m)}=\hat{w}_{t}^{(m)}/\sum\limits_{m'}{\hat{w}_{t}^{(m')}}$ (for $m=1:N_p$)
\State Compute estimates: $\hat{\mathbf{x}}_{t}=\sum\limits_{m}\hat{w}_{t}^{(m)}\mathbf{x}_{t}^{(m)}$
\State Resample with replacement from $\{\hat{w}_{t}^{(m)},\mathbf{x}_{t}^{(m)}\}^{N_p}_{m=1}$
\end{algorithmic}
\end{algorithm}

\subsection{Distributed Particle Filtering}\label{subsec:dpf}
For a distributed implementation of the PF, we want to avoid exchanging observations, while at the same time maintaining a common set of samples and weights at every time step. If we can guarantee that the samples and weights at time $t-1$ are common, then common samples\footnote{Although different sample realizations is advantageous in many applications, this is not the case in DPF since one set of particles is sufficient for one target.} at time $t$ can be achieved by providing all nodes with the same seed for random number generation (i.e., by ensuring that their pseudo-random generators are in the same state at all times). Ensuring common weights for all nodes can be achieved by means of a belief consensus (BC) algorithm. BC formally aims to compute, in a distributed fashion the product of a number of real-valued functions over the same variable
\be
\mathrm{BC}(f_1(x),f_2(x),\ldots,f_{N_s}(x))=\prod_{n=1}^{N_s}f_n(x).
\ee
However, most BC algorithms are not capable to achieve exact consensus in a finite number of iterations (except BP consensus in tree-like graphs; see Section \ref{subsec:bc}). As we require \emph{exact} consensus on the weights, we additionally apply max-consensus\footnote{Min-consensus or average over min- and max-consensus can be also applied \cite{Farahmand2011}.} \cite{Ustebay2011,Olfati-Saber2004},  
\be
\mathrm{MC}(f_1(x),f_2(x),\ldots,f_{N_s}(x))=\max_{n}f_n(x),
\ee
which computes the exact maximum over all arguments 
in a finite number of iterations (equal to the diameter of the graph). In particular, max-consensus  can be implemented as follows: denoting the function at iteration $i$ at node $n$ by $f^{(i)}_n(x)$ (with $f^{(0)}_n(x)= f_n(x)$) every node executes the following rule:
\be
f^{(i)}_n(x) = \max \left(f^{(i-1)}_n(x),~\max_{u \in G_n} f^{(i-1)}_u(x)\right).
\ee
Note that max-consensus is applied before computing the estimates in order to avoid disagreement of the estimates over network. This approach, already used in \cite{Ustebay2011,Farahmand2011}, can be considered as a consensus-based approach to synchronize the particles for the next time instant. In addition, we assume that time-slot synchronization is performed in distributed way using any appropriate standard technique, see \cite{Leng2011} and references therein.

The final algorithm is shown in Alg.~\ref{dist-SIR}. Observe that DPFs operate on two distinct time scales: a slow time scale (related to time slots) in which the target moves, and the observations are taken, and a fast time scale (related to iterations) in which BC/MC is executed for a specific time slot. Note also that in comparison with CPF, DPF has the following advantages: i) the energy consumption is balanced across the network; ii) reduced communication cost in large-scale networks; iii) every node has access to the posterior distribution; and iv) no knowledge required of the locations, observations, or observation models of any other node.

\subsection{Belief Consensus Algorithms}\label{subsec:bc}

 Motivated by their scalability and robustness to failures \cite{Olfati-saber2006,Aysal2009, Dimakis2010, Crick2003}, we consider five variants of BC: standard BC (SBC) \cite{Olfati-saber2006}, randomized gossip (RG) \cite{Boyd2006, Dimakis2010}, Metropolis BC (MBC) \cite{Xiao2004}, broadcast gossip (BG) \cite{Aysal2009, Dimakis2010}, and belief propagation (BP)  \cite{Crick2003,Pearl1988} consensus. While BP consensus has not been applied within DPF, the other four BCs have already been applied \cite{Ustebay2011,Farahmand2011,Hlinka2012,Liu2009}.

We will now describe these five distinct BC algorithms corresponding to line 3 in Alg.~\ref{dist-SIR}. Although we use BC to perform the consensus on particle weights, the algorithms will be presented in general form (with continuous variables) in order to relate them with the state-of-the-art algorithms.

\subsubsection{Standard BC}\label{subsubsec:sbc}
Standard BC (SBC) \cite{Olfati-saber2006} is defined in following iterative form:
\be
\label{bc-std}
M_n^{(i)} (\mathbf{x}_t) = M_n^{(i-1)} (\mathbf{x}_t) \prod\limits_{u \in G_n }{\left(\frac{{M_u^{(i - 1)} (\mathbf{x}_t)}}{{M_n^{(i - 1)} (\mathbf{x}_t)}}\right)^{\xi}},
\ee
where 
$M_n^{(i)}(\mathbf{x}_t)$ represents the approximation at iteration $i$ of the global likelihood of the variable $\mathbf{x}_t$, and $\xi$ is update rate, which depends on maximum node degree in the network ($\eta_{\max}=\max_n{\left|G_n\right|}$).\footnote{Note that the logarithm of (\ref{bc-std}) corresponds to standard average consensus algorithm \cite[eq. (4)]{Olfati-saber2006}. All computations are done in the log-domain to avoid numerical problems.} The update rate $\xi \approx 1/\eta_{\max}$ provides sufficiently fast convergence for the constant weight model \cite{Olfati-Saber2004, Xiao2004}. Optimized constant weights \cite{Xiao2007} can be found using distributed convex optimization \cite{Johansson2008,Aoyama2010}, but the optimization is generally complex since the Laplacian matrix has to be estimated at each node.
We initialize (\ref{bc-std}) by 
\be
\label{eq:sbc-init}
M_{n}^{(1)}({\mathbf{x}_t}) =p({y_{n,t}}|{\mathbf{x}_t}).
\ee
Convergence is guaranteed for all connected graphs in a sense that  \cite{Olfati-saber2006, Xiao2004}
\be
\label{bc-conv}
\mathop {\lim }\limits_{i \to \infty } M_n^{(i)}(\mathbf{x}_t) =  \left({{\prod\limits_{n' \in G_t} p({y_{n',t}}|{\mathbf{x}_t})}}\right)^{1/N_s},
\ee
from which the desired quantity, $\prod_{n \in G_t} p({y_{n,t}}|{\mathbf{x}_t})$, can easily be found, for any value of $\mathbf{x}_t\in \{ \mathbf{x}_t^{(1)},\ldots,\mathbf{x}_t^{(N_p)}\}$. However, in practical circumstances, we run SBC a finite number of iterations ($i=1,2,\ldots,N_i^{\rm{SBC}}$), so the result will be an approximation of the real likelihood.

If the maximum node degree ($\eta_{\max}$) and number of nodes ($N_s$) are not known a priori, we need to estimate them in distributed way.\footnote{\red{While any upper bound on $\eta_{\max}$ guarantees convergence, estimation of $\eta_{\max}$ is preferable to increase convergence speed of SBC.}} The estimation of maximum node degree can be  done using max-consensus, while $N_s$ can be determined \cite{Pham2009} by setting an initial state of one node to 1, and all others to 0. By using average consensus \cite{Olfati-Saber2004}, all nodes can obtain the result $1/N_s$, i.e., the inverse of the number of nodes in the network.	

SBC was used for consensus on weights in \cite{Farahmand2011}. Moreover, there are a number of specific instances of SBC (e.g., \cite{Gu2007a,Gu2008,Hlinka2011}) that represent the beliefs in parametric form (e.g., Gaussian, or Gaussian mixture), and make consensus on their parameters.
~\\
\subsubsection{BC based on Randomized Gossip}\label{subsubsec:rg}
Gossip-based algorithms \cite{Dimakis2010} can be also used to achieve consensus in a scalable and robust way. We consider randomized gossip (RG) \cite{Boyd2006}. In RG, it is assumed that all the nodes have internal clocks that tick independently according to a rate of e.g., a Poisson process \cite{Dimakis2010}. When the clock of the $n$-th node ticks, node $n$ and one of its neighbors (randomly chosen) exchange their current estimates, and make the update. In case of BC based on RG, we need to achieve convergence to the geometrical average, so at the $i$-th clock tick of node $n$, the nodes $n$ and $u$ make the following operation:
\be
\label{eq:rg-belief}
M_u^{(i)} (\mathbf{x}_t)=M_n^{(i)} (\mathbf{x}_t)=\left(M_u^{(i-1)} (\mathbf{x}_t) M_n^{(i-1)} (\mathbf{x}_t)\right)^{1/2}
\ee
where $u \in {G_n}$, and all other nodes $r$ in the network ($r \notin\{n,u\}$) do not make any update (i.e., $M_r^{(i)} (\mathbf{x}_t)=M_r^{(i-1)} (\mathbf{x}_t)$). Initialization is done using (\ref{eq:sbc-init}). In order to have the same communication cost as SBC, RG should run approximately $N_i^{\rm{RG}}=\lceil N_i^{\rm{SBC}}  N_s/2\rceil$ iterations. Finally, we again need to estimate $N_s$ using the same approach as for SBC.

RG has been  used in \cite{Ustebay2011} for consensus on weights, and a specific instance (with Gaussian approximation) in \cite{Oreshkin2010}.
~\\
\subsubsection{BC based on Broadcast Gossip}\label{subsubsec:bg}
The main problem of RG is that once a node broadcasts data, only one of its neighbors performs an update. It is natural to expect that if all the neighbors perform an update, the convergence will be faster. To address this problem, broadcast gossip (BG) has been proposed \cite{Aysal2009}, in which a randomly chosen node broadcasts its current estimate, and all of its neighbors (within communication radius $R$) perform an update. It has been shown \cite{Aysal2009} that, \red{due to its asymmetric nature,} BG converges only in expectation to the real average value.\footnote{\red{Note that the underlying communication graph is still undirected, since all the nodes have to be capable to broadcast their estimates (even if they don't do so at each iteration).}} 

In our case, we need to achieve convergence to the geometrical average (\ref{bc-conv}), so at the $i$-th clock tick of node $n$ all the nodes make the following operation:
\be
\label{eq:bg-belief}
M_u^{(i)} (\mathbf{x}_t)= \left\{ \begin{array}{l}
M_u^{(i-1)} (\mathbf{x}_t)^\gamma M_n^{(i-1)} (\mathbf{x}_t)^{1-\gamma},\hfill~\,u \in {G_n}\\
M_u^{(i-1)} (\mathbf{x}_t),\hfill~\mathrm{otherwise.}
\end{array} \right.
\ee
where $0<\gamma<1$ is the mixing parameter. Again, initialization is done using (\ref{eq:sbc-init}). To synchronize communication cost with previous BC methods, we run BG $N_i^{\rm{BG}}=N_i^{\rm{SBC}} N_s$ iterations. It is again necessary to apply average consensus to estimate $N_s$.

A variant of this method, with particle compression based on support vector machine, has been already applied in \cite{Liu2009}.

\emph{Comment:} Regarding the choice of  $\gamma$, it has been shown in \cite{Aysal2009} that its optimal value depends on the algebraic connectivity of the graph, which is the second smallest eigenvalue of the Laplacian matrix \cite{Aysal2009,Olfati-Saber2004}. However, this parameter is not available in the distributed scenario, so an empirical study has been used \cite{Aysal2009} to find the optimal value of $\gamma$. Therefore, we will model $\gamma$ as a function of average node degree $\bar{\eta}$ in the network, since $\bar{\eta}$ can be easily estimated using average consensus. We found that an optimal $\gamma$ can be modeled as
\be
\label{eq:gamma}
\gamma(\bar{\eta})=1 - a{e^{ - b\bar{\eta}}} \in (0,1),
\ee
with parameters $a$ and $b$, which can be estimated by calibration.
~\\
\subsubsection{Metropolis BC}\label{subsubsec:mbc}
An important problem of the SBC method is that it uses a constant weight model, i.e., $\xi_{nu}=\xi$ for each link $(n,u)$ in the graph, which will not provide good performance in asymmetrical graphs. Instead of this model, we can use so-called Metropolis weights, which should provide  faster convergence \cite{Xiao2004}. For our problem, this leads to Metropolis BC (MBC), with the following update rule:
\be
\label{metro-bc}
M_n^{(i)} (\mathbf{x}_t) = M_n^{(i-1)} (\mathbf{x}_t)^{\xi_{nn}} \prod\limits_{u \in G_n} M_u^{(i-1)} (\mathbf{x}_t)^{\xi_{nu}},
\ee
where the weight on an link $\{n,u\}$ is given by:
\be
\label{metro-wei}
{\xi _{nu}} = {\xi _{un}} = \left\{ {\begin{array}{*{20}{l}}
{1/\max ({\eta _n},{\eta _u}),\,\,\,\,{\rm{for}}\,\,u \ne n}\\
{1 - \sum\nolimits_{u' \in {G_n}} {{\xi _{u'n}}} ,\,\,\,\,\,{\rm{for}}\,\,u = n.}
\end{array}} \right.
\ee
The initialization is the same as for other BC methods, and the number of iterations is the same as for SBC ($N_i^{\rm{MBC}}=N_i^{\rm{SBC}}$). This approach also guarantees convergence \cite{Xiao2004} to (\ref{bc-conv}).\footnote{Provided that the graph is not bipartite. Otherwise, $\max ({\eta _n},{\eta _u})$ should be replaced with $\max ({\eta _n},{\eta _u})+1$.} 
As we can see, this method is more suitable than SBC for distributed implementation, since a node needs to know only the local degrees of its neighbors. However, $N_s$ still has to be estimated.

A specific instance of MBC, in which the beliefs belong to the exponential family, is used in \cite{Hlinka2012}.
~\\
\subsubsection{BC based on Belief Propagation}\label{subsubsec:bp}

\begin{figure}[!t]
\centering
\includegraphics[width=0.4\textwidth]{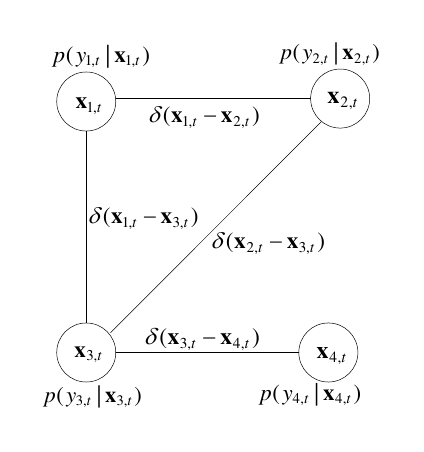}
\caption{Example of a graphical model for BP consensus, for a network with 4 sensors.}
\label{fig:graph-model}
\end{figure}

Belief propagation (BP) \cite{Crick2003,Pearl1988} is a way of organizing the global computation of marginal beliefs in terms of smaller local computations within the graph. To adapt it for BC, we define the following function:
\be
f(\mathbf{x}_{1,t},\mathbf{x}_{2,t},\ldots,\mathbf{x}_{N_s,t})=\prod\limits_{n} p({y_{n,t}}|{\mathbf{x}_{n,t}})\prod_{u \in G_n} \delta({\mathbf{x}_{n,t}-\mathbf{x}_{u,t}}).
\ee
 Running BP on the corresponding graphical model (see example in Figure \ref{fig:graph-model}) yields the marginals $M_n(\mathbf{x}_{n,t})$ of the function $f(\mathbf{x}_{1,t},\mathbf{x}_{2,t},\ldots,\mathbf{x}_{N_s,t})$. It is easily verified that for every $n$
\begin{eqnarray}
\label{marginals1}
M_n(\mathbf{x}_{n,t})  &= &  \sum_{\mathbf{x}_{u \neq n,t}} f(\mathbf{x}_{1,t},\mathbf{x}_{2,t},\ldots,\mathbf{x}_{N_s,t})\\
& =  &  \prod_{n'}p({y_{n',t}}|{\mathbf{x}_{n,t}}).\label{marginals1}
\end{eqnarray}

The BP message passing equations are now as follows: the belief at iteration $i$ is given by \cite[eq.~(8)]{Ihler2005a}
\be
\label{beliefs1}
M_n^{(i)}(\mathbf{x}_{n,t} ) \propto p({y_{n,t}}|{\mathbf{x}_{n,t}}) \prod\limits_{u \in G_n } {m_{un}^{(i)} (\mathbf{x}_{n,t} )},
\ee
while the message from node $u\in G_n$ to node $n$ is given by \cite[eq.~(9)]{Ihler2005a}
\begin{eqnarray}
\label{msgs1}
m_{un}^{(i)} (\mathbf{x}_{n,t} )  &\propto &  \int\limits_{\mathbf{x}_{u,t} } {\delta({\mathbf{x}_{n,t}-\mathbf{x}_{u,t}})\,\frac{{M_u^{(i-1)} (\mathbf{x}_{u,t} )}}{{m_{nu}^{(i-1)} (\mathbf{x}_{u,t} )}}}d\mathbf{x}_{u,t}\\
& =  &\frac{{M_u^{(i - 1)} (\mathbf{x}_{n,t})}}{{m_{nu}^{(i - 1)} (\mathbf{x}_{n,t})}}.\label{msgs2short}
\end{eqnarray}
We note that since all variables are the same, we can write $\mathbf{x}_{n,t}=\mathbf{x}_{u,t}=\mathbf{x}_t$. Some manipulation yields (see \ref{subsec:bp-eq})
\be
\label{new-bc}
M_n^{(i)} (\mathbf{x}_t) \propto M_n^{(i-2)} (\mathbf{x}_t) \prod\limits_{u \in G_n }{\left(\frac{{M_u^{(i - 1)} (\mathbf{x}_t)}}{{M_n^{(i - 2)} (\mathbf{x}_t)}}\right)},
\ee
which represents the consensus algorithm based on BP. Since BP consensus uses the same protocol as SBC and MBC, we should run it $N_i^{\rm{BP}}=N_i^{\rm{SBC}}$ iterations. This method is initialized by $M_n^{(1)}(\mathbf{x}_t)=p({y_{n,t}}|{\mathbf{x}_{t}})$. We also need to set $M_n^{(2)}(\mathbf{x}_t)$ in order to run the algorithm defined by (\ref{new-bc}). Using (\ref{beliefs1}) and (\ref{msgs1}), and assuming that $m_{nu}^{(1)} (\mathbf{x}_{t})=1$, we find
\be
\label{semi-local-init}
M_n^{(2)}(\mathbf{x}_t)=p({y_{n,t}}|{\mathbf{x}_{t}})\prod_{u \in G_n }p({y_{u,t}}|{\mathbf{x}_{t}}).
\ee

BP consensus (as a specific instance of BP) guarantees convergence to $C\prod_{n}p({y_{n,t}}|{\mathbf{x}_{t}})$ for cycle-free network graphs  \cite{Pearl1988,Yedidia2003, Weiss1998}, where $C$ is an irrelevant normalization constant.\footnote{It is irrelevant since the weights in Alg. \ref{dist-SIR} will be normalized later anyway.} When the network has cycles, the beliefs are only approximations of the true marginals given by \eqref{marginals1} (more details in \ref{subsec:bp-corr}). Comparing BP consensus with previous consensus methods, we can see that BP-consensus agrees on product of all local evidences (not the $N_s$-th root of the product), and does not rely on knowledge of any other parameters. Therefore, it is more robust to the changes in the network.

\subsection{Communication Cost Analysis}\label{subsec:comm-cost}

In this section, we analyze the communication cost of the DPF methods, and compare with the cost of NCPF. We denote by $N_{\mathrm{pack}}$  the number of packets that any node $n$ broadcasts at any time $t$. We assume that one packet can contain $P$ scalar values (mapping from scalars to bits is not considered). In most hardware platforms, $P\gg1$, and the energy required to transmit one packet does not significantly depend on the amount of data within it. In this analysis, we neglect the cost of determining $\eta_{\max}$, $\bar{\eta}$, and $N_s$ and the cost of time-slot synchronization. Consequently, all DPF methods will have the same communication cost.

\subsubsection{Cost of DPF}
At every iteration (except the first), nodes  transmit $N_p$ weights. In addition, nodes must perform MC, which also requires transmission of the weights in each iteration. The number of iterations of the BC\footnote{We use $N_i^{\rm{SBC}}$ to count iterations ($N_{\mathrm{it}}=N_i^{\rm{SBC}}$). All other DPF methods run the number of iterations which ensures the same communication cost as explained in Section \ref{subsec:bc}.} is $N_{\mathrm{it}}$. The number of iterations of the MC is equal to the diameter of the graph $D_g$. Thus, the average cost of DPF per node and per time slot  is 
\be
\label{eq:cost-dpf}
N_{\mathrm{pack}}^{\mathrm{DPF}}\approx \left\lceil \frac{N_p}{P} \right\rceil (D_g+N_{\mathrm{it}}-1).
\ee
We see that the DPF methods are fully scalable, since increasing the number of the nodes in a fixed deployment area will not significantly affect the cost. Although beyond the scope of  this paper, we mention that if one prefers to use parametric approximations \cite{Gu2007a,Gu2008,Hlinka2011,Hlinka2012} instead of consensus on weights, only parameters of the beliefs should be transmitted in each iteration.

\subsubsection{Cost of NCPF}
In contrast to DPF, NCPF does not require transmission of the weights, but only the local data as described in Section \ref{subsec:ncpf}. We denote the number of these scalar values as $N_{\mathrm{data}}$. The amount of data will accumulate with iterations since the node has to transmit its own data and all received data. Since the number of iterations is equal to $D_g$, the cost can be approximated by:
\be
\label{eq:cost-ncpf}
N_{\mathrm{pack}}^{\mathrm{NCPF}} \approx \sum\limits_{k = 0}^{{D_g} - 1} \left\lceil \frac{\bar{\eta}^{k}N_{\mathrm{data}}}{P} \right\rceil,
\ee
where we approximate the degree of the each node with average network degree ($\bar{\eta}$),  and $\sum_{i = 0}^{k} \bar{\eta}^{i}$Ê by $\bar{\eta}^{k}$.

\subsubsection{Comparison between DPF and NCPF}
Making the reasonable assumption that  $N_{\mathrm{it}} = D_g+1$, we can quantify when DPF is preferred over NCPF, i.e., when  $N_{\mathrm{pack}}^{\mathrm{DPF}}<N_{\mathrm{pack}}^{\mathrm{NCPF}}$:
\red{
\be
\label{eq:cost-ncpf-dpf}
\left\lceil {\frac{{{N_p}}}{P}} \right\rceil  < \frac{1}{{2{D_g}}}\sum\limits_{k = 0}^{{D_g} - 1} {\left\lceil {\frac{{{{\bar \eta }^k}{N_{\mathrm{data}}}}}{P}} \right\rceil }.
\ee
}
This condition is important in order to avoid over-using of consensus-based methods. For example, if the network is fully-connected ($D_g=1$), or if the packet size is sufficiently large to afford transmission of all accumulated data (i.e., $P>{{{\bar \eta }^{D_g-1}}N_{\mathrm{data}}}$), NCPF should be applied. On the other hand, if the communication radius is very small (i.e., if $D_g$ is very large), DPF methods should be applied. 

A similar comparison can be done with CPF, but its cost depends on many factors \cite{Hlinka2013}, including the routing protocol, and the position of the fusion center. In general, the cost of CPF is much smaller than NCPF since there is only one (instead of $N_s$) fusion center. However, even with reduced cost, CPF is not a good alternative due to the uneven energy consumption over nodes, and since the posterior cannot be accessed from any node (see also Section \ref{subsec:cpf}).

Regarding computational complexity, DPF methods require $\mathcal{O}(\left|G_n\right|N_{it}N_p)$ operations per node at time $t$, while NCPF requires only $\mathcal{O}(\left|G_t\right|N_p)$ operations per node. However, the energy consumption in typical hardware platforms is mainly caused by communication, while internal computation is very cheap \cite{Patwari2005}. Therefore, DPF methods are still preferred over NCPF.

\section{Simulation Results}\label{sec:sim}

\begin{figure*}[!t]
\centerline{
\subfloat[]{\includegraphics[width=0.4\textwidth]{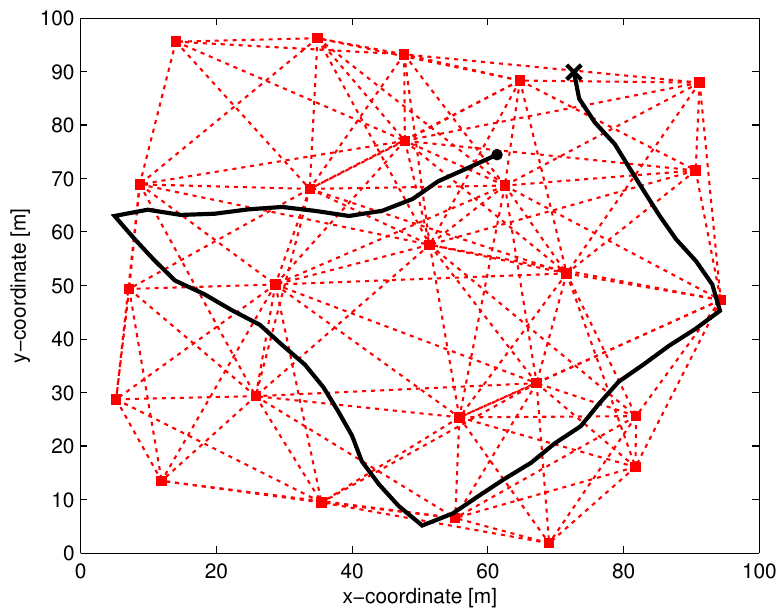}%
\label{fig:trackTrueSemi}}
\hfil
\subfloat[]{\includegraphics[width=0.4\textwidth]{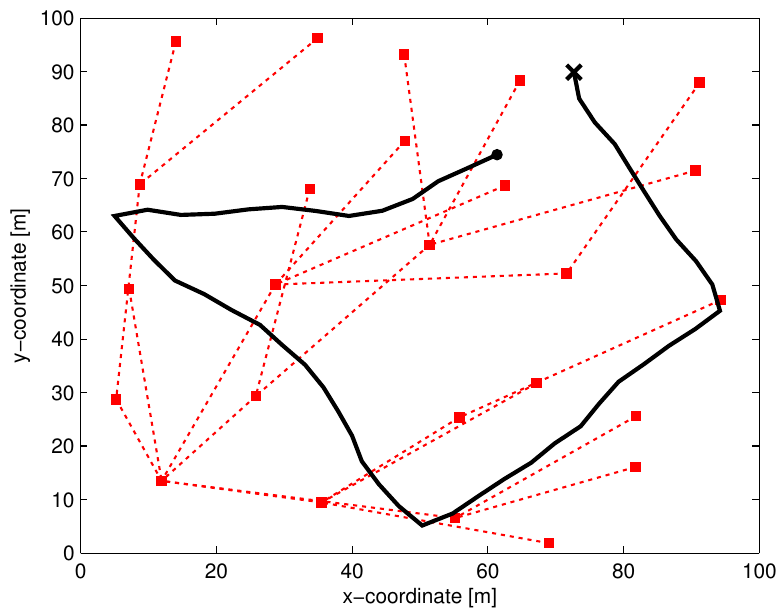}%
\label{fig:trackTrueTree}}
}
\caption{Illustrations of the target track in: (a) semi-random, and (b) tree network, for $R=45$ m. The start point of the track is marked with a small black disc, and the end point with an X. Sensors are marked with the squares, and communication links with dashed lines.}
\label{fig:trackTrue}
\end{figure*}

\subsection{Simulation Setup and Performance Metrics}

We assume that there are $N_s=25$ sensors semi-randomly\footnote{Semi-random network is created by adding jitter to a regular 2D grid.} deployed in a 100m x 100m area. The positions of these sensors are perfectly known. There is also one target in the area, with state ${\mathbf{x}_t} = {[{x_{1,t}}\,\,{x_{2,t}}\,\,{\dot x_{1,t}}\,\,{\dot x_{2,t}}]^T}$, where $x_{1,t}$ and $x_{2,t}$ represent 2D position of the target, and ${\dot x_{1,t}}$ and ${\dot x_{2,t}}$ the 2D velocity of the target. The target moves with constant amplitude of the speed of $5$m/s, according to Gaussian random walk, given by 
\be
\label{eq:motion}
{\mathbf{x}_{t + 1}} = \left[ {\begin{array}{*{20}c}
   \mathbf{I}_2 & {{T}_{s} \mathbf{I}_2 }  \\
   {{\textbf{0}}_{2} } & \mathbf{I}_2  \\
\end{array}} \right]\mathbf{x}_t + \left[\begin{array}{l}
 \frac{{{T}_{s} ^{2} }}{{2}}\mathbf{I}_2  \\ 
 {T}_{s} \mathbf{I}_2  \\ 
 \end{array} \right] \mathbf{u}_t,
\ee
where $T_s$ is the sampling interval (here set to 1 second), and $\mathbf{I}_2$ and $\textbf{0}_{2}$ represent the identity and zero 2 $\times$ 2 matrices, respectively. The process noise $\mathbf{u}_t$ is distributed according to zero-mean Gaussian with covariance matrix $0.5\mathbf{I}_2~(\rm{m/s^2})^2$. The target is tracked for $N_t=50$ seconds. 

We set the sensing radius to $r=25$m, and consider two values of communication radius: $R=25$m and $R=45$m (corresponding to $\bar{\eta}=3.08$ and $\bar{\eta}=9.44$, respectively).  In addition, we consider a tree configuration created as a spanning tree\footnote{Tree configurations should not be created using an online algorithm since it would require a routing of data. We simply assumed that the tree configuration is established offline or is an inherent property of the network.} from semi-random network. An example of a target track in a loopy and tree network is shown in Figure \ref{fig:trackTrue}. The inclusion of tree topologies is motivated by certain target tracking applications, such as tunnels.%

We assume that the observations are distances to the target, i.e., for $n \in G_t$,
\be
\label{eq:measurement}
{y_{n,t}} = g_n(\mathbf{x}_t)=\left\| {\mathbf{l}_n  - \left[ {x_{1,t} \,\,x_{2,t} } \right]^T } \right\| + {v_{n,t}}.
\ee
The observation noise $v_{n,t}$ is distributed according to Gaussian mixture with two components, a typical model in presence of non-line-of-sight signals. The parameters of this noise are set to following values: means $(1\rm{m},10\rm{m})$, variances $(1\rm{m^2},1\rm{m^2})$ and mixture weights $(0.9,0.1)$.

 For mixing parameter in BG, given by (\ref{eq:gamma}), we set $a=0.49$, and $b=0.17$ found by calibration.  We use $N_p=500$ particles. The results are averaged over $N_{R}=500$ Monte Carlo runs. In each run, we generate different network, track, observations and particle seed.

\begin{figure*}[!t]
\centerline{
\subfloat[]{\includegraphics[width=0.45\textwidth]{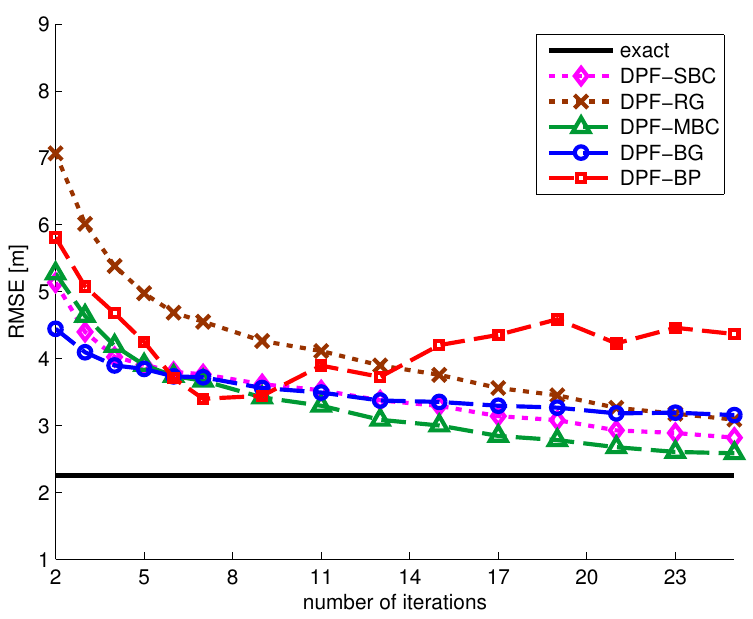}
}
\hfil
\subfloat[]{\includegraphics[width=0.453\textwidth]{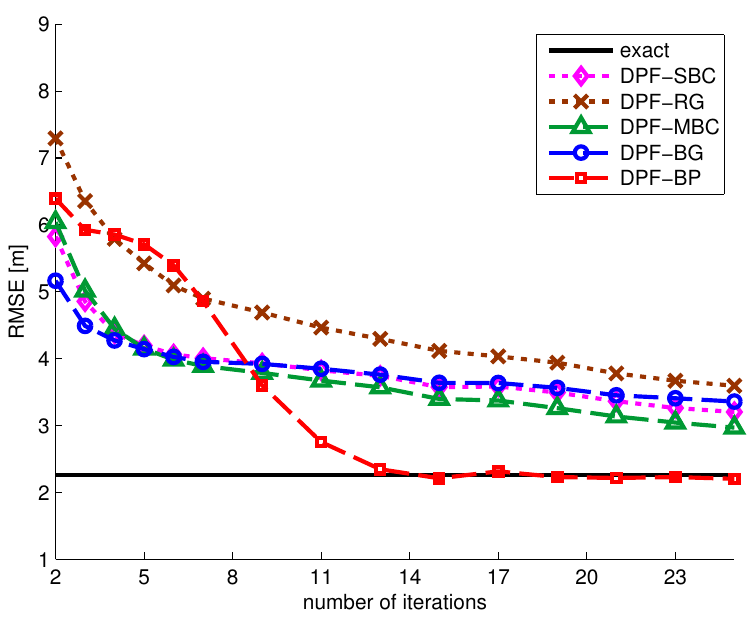}}
}
\centerline{
\subfloat[]{\includegraphics[width=0.45\textwidth]{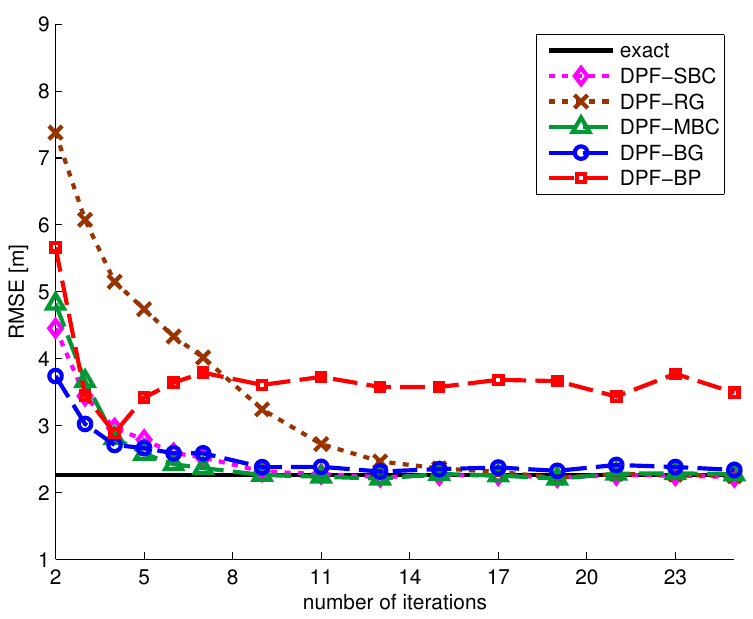}}
\hfil
\subfloat[]{\includegraphics[width=0.45\textwidth]{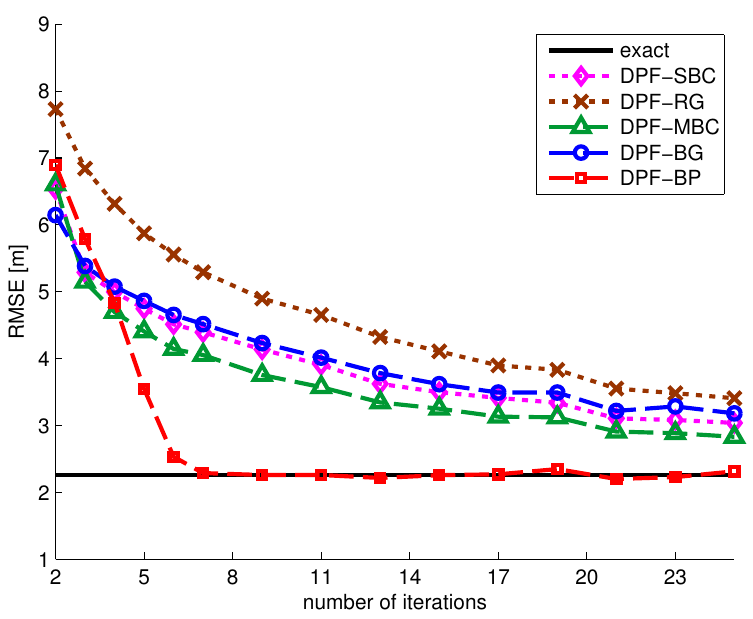}}
}
\caption{ RMSE of DPF methods as a function of the number of iterations for  (a) semi-random ($R=25$ m),  (b) tree ($R=25$ m), (c) semi-random ($R=45$m), and (d) tree ($R=45$m).}
\label{fig:PFiter25nodes}
\end{figure*}

We will compare five DPF methods (DPF-SBC, DPF-MBC, DPF-RG, DPF-BG, and DPF-BP). Moreover, as a benchmark, we show the performance of the exact approach, which corresponds to performance of CPF and NCPF. We consider root-mean-square error (RMSE) in the position error \red{$e_{\mathrm{rms},t}$ and $e_{\mathrm{rms}}$} as a performance metrics. Introducing $e_{n,t,s}$ as the target positioning error \red{(i.e., Euclidean distance between the true and estimated position of the target)} at node \emph{n}, at time $t$ in simulation run $s$, we have
\red{
\be
\label{eq:rmse}
{e_{\mathrm{rms},t}} = \sqrt {\frac{\sum_{n=1}^{N_s} \sum_{s=1}^{N_R} {e_{n,t,s}^2} } {N_s N_{R}}},~~~{e_{\mathrm{rms}}} = \sqrt{\frac{\sum_{t=1}^{N_t} e_{\mathrm{rms},t}^2}{N_t}}
\ee
}

\begin{figure*}[!t]
\centerline{
\subfloat[]{\includegraphics[width=0.45\textwidth]{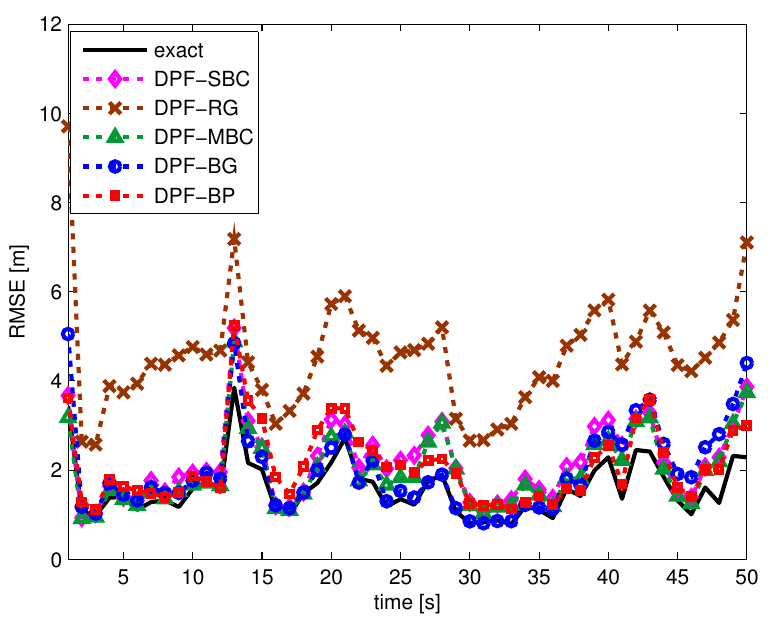}%
\label{fig:rmseTime45mSemi}}
\hfil
\subfloat[]{\includegraphics[width=0.45\textwidth]{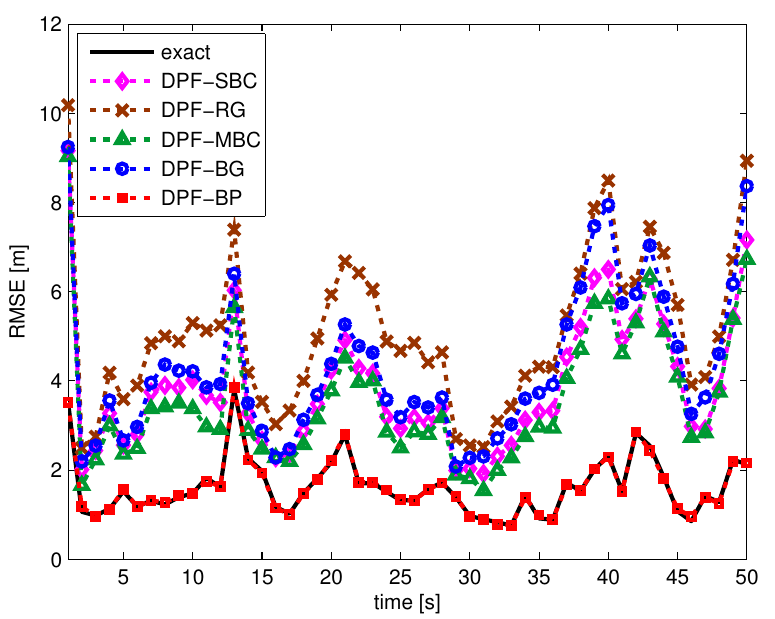}%
\label{fig:rmseTime45mTree}}
}
\caption{RMSE of DPF methods as a function of time for: (a) semi-random network from Figure \ref{fig:trackTrueSemi} ($N_{\mathrm{it}}=4$), and (b) tree network from Figure \ref{fig:trackTrueTree}, ($N_{\mathrm{it}}=7$).}
\label{fig:rmseTime45m}
\end{figure*}

\subsection{Performance Results}

We first investigate the RMSE \red{$e_{\mathrm{rms}}$} as a function of the number of iterations $N_{\mathrm{it}}$, for different communication radii and network topologies. The results are shown in Figure \ref{fig:PFiter25nodes}. We can draw a number of conclusions. As expected, exact methods provide the best RMSE performance, as they compute the exact likelihood corresponding to (\ref{eq:g-lhood}). Secondly, all DPF methods (except for DPF-BP) are capable to reach asymptotically the performance of the exact approach. Among the DPF methods in the semi-random topology (Figures \ref{fig:PFiter25nodes}a and \ref{fig:PFiter25nodes}c), DPF-MBC provides the fastest convergence, for both considered communication radii, \red{though DPF-BG and DPF-SBC achieve similar performance for $R=45$ m. } DPF-BP achieves a minimal RMSE for $N_{\mathrm{it}} = D_g+1$, since then all local likelihoods are available at each node (see also \ref{subsec:bp-corr}). 
A further increase of the number of iterations will only increase the amount of over-counting of the local likelihoods, thus leading to biased beliefs. In the tree topology (Figures \ref{fig:PFiter25nodes}b and \ref{fig:PFiter25nodes}d), the situation is very different: DPF-BP provides the fastest convergence and \textit{exact} result after  $D_g+1$ iterations. However, note that $D_g$ is typically higher for tree networks than loopy networks. Finally, we also see that: i) increasing $R$ increases the convergence speed of all DPFs, and ii) DPF-RG performs the worst for both considered radii and configurations.

Secondly, in Figure \ref{fig:rmseTime45m}, we analyze the RMSE  \red{$e_{\mathrm{rms},t}$}  as a function of time for the track and the networks from Figure \ref{fig:trackTrue}. We consider the case in which $R=45$ m, and $N_{\mathrm{it}}$ is set to provide the optimal performance of DPF-BP. Here, in each Monte Carlo run, we generate different observations and particle seeds, but keep the same track and the network. In case of semi-random network (Figure \ref{fig:trackTrueSemi}), we see that all DPFs (except DPF-RG) provide the performance close to the exact approach. On the other hand, in the tree network, DPF-BP performance overlaps with the exact approach. In general, DPF-BP can be used in all configurations, if we know (or are able to estimate) $D_g$. However, in practice, the network may have no knowledge of $D_g$, so DPF-BP would lead to biased results.

\begin{figure*}[!t]
\centerline{
\subfloat[]{\includegraphics[width=0.33\textwidth]{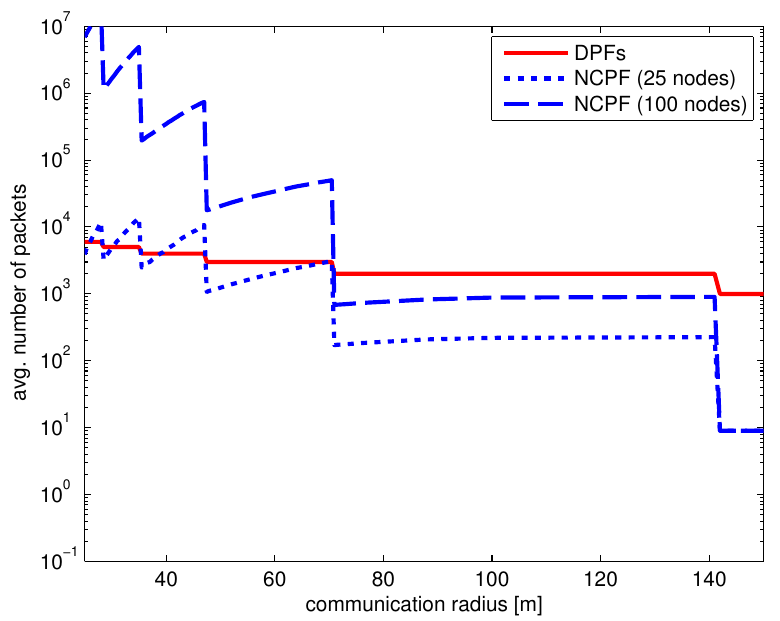}\label{fig:msgCommP1}}
\hfil
\subfloat[]{\includegraphics[width=0.33\textwidth]{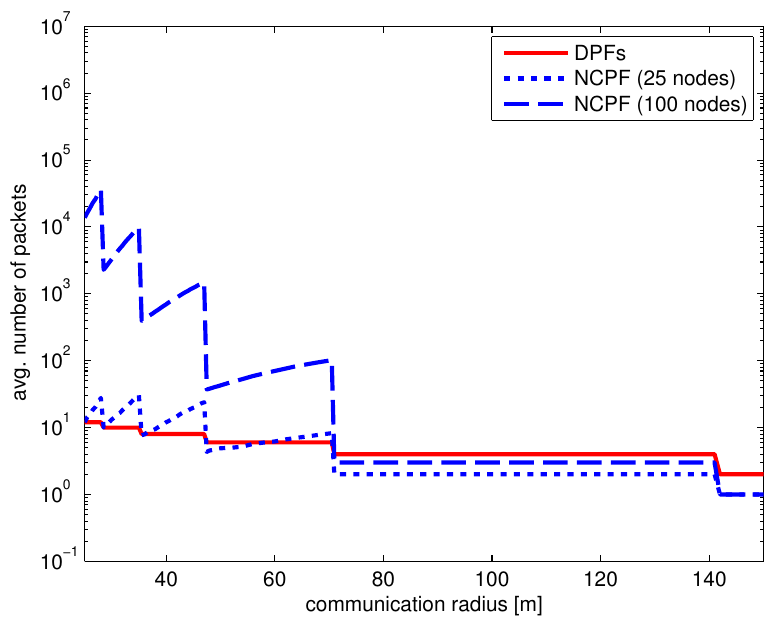}\label{fig:msgCommP500}} 
\hfil
\subfloat[]{\includegraphics[width=0.33\textwidth]{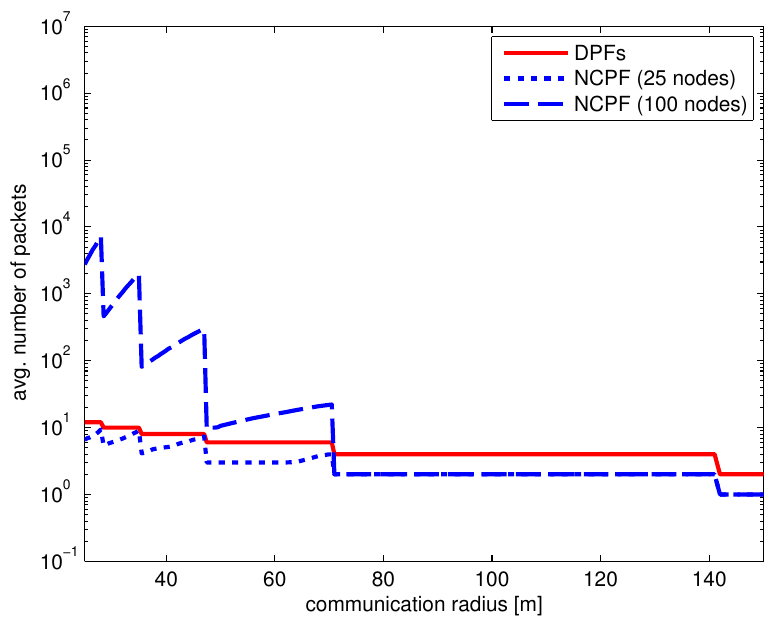}\label{fig:msgCommP2500}}
}
\caption{ Communication cost comparison as a function of the communication radius, for: (a) $P=1$, (b) $P=N_p$, and (c) $P=5N_p$.}
\label{fig:PFmsgComm}
\end{figure*}

Finally, we evaluate the communication cost per node, by analyzing the average number of packets per node as a function of the communication radius $R$ for $N_{\mathrm{it}}=\left\lceil L/R \right\rceil+1$. Here $L$ is the diameter of the deployment area ($L=100\sqrt{2}$ m, in our case). We consider networks with 25 and 100 nodes, and packet sizes of $P=1$, $P=N_p$, and $P=5N_p$, where $N_p=500$, and $N_{\mathrm{data}}=9$ (i.e., it includes 2 scalars for the position, 6 scalars for the observation model, and 1 scalar for the observation). To count the number of packets, we simulated the degree of each node in all networks, and applied equations (\ref{eq:cost-dpf})--(\ref{eq:cost-ncpf}). As we can see from Figure \ref{fig:PFmsgComm}, DPF-based methods provide nearly constant communication cost as a function of $R$, since (\ref{eq:cost-dpf}) only  depends linearly on $D_g$, and it does not depend on $N_s$. Thus, these methods are scalable. On the other hand, the communication cost of NCPF is highly sensitive to $R$ and $N_s$: the cost increases as $R$ increases (while $\left\lceil L/R \right\rceil$ is fixed), and decreases significantly when $\left\lceil L/R \right\rceil$ decrements its value (e.g., for $R=50\sqrt{2}$). Overall, decreasing $\left\lceil L/R \right\rceil$ has the largest effect (see (\ref{eq:cost-ncpf})), so the total cost has a decreasing tendency with $R$. In addition, since the increased $N_s$ affects $\bar{\eta}$, the communication cost will be significantly larger. Regarding the effect of \emph{P}, we can see that larger values of \emph{P} will make NCPF cheaper, as more data can be aggregated in one packet. Finally, comparing with NCPF, we can see that DPF  methods have a lower communication cost for $R<70$m, except when $P$ is very large (as in Figure \ref{fig:msgCommP2500}).

\subsection{Discussion}
According to previous results, we can see that BC-based DPF methods should be applied only in sparse networks, i.e., in networks with relatively high diameter. Otherwise, NCPF may be used since it is cheaper to flood a few scalars instead of iteratively broadcasting weights or parameters of the likelihood function. Moreover, we found that DPF-MBC is the fastest in loopy graphs, \red{closely followed by DPF-SBC, and DPF-BG.} Hence, DPF-MBC is preferred in unconstrained areas in which the loops are expected, such as conference halls, airports, and warehouses \cite{Savic2011a}. On the other hand, DPF-BP is only suitable for confined areas, in which long chains of sensors, globally forming a tree, are deployed. Typical examples are subway, roadway and mine tunnels \cite{Savic2013fusion,Chehri2012}. Note that the robustness is decreased in tree graphs since a node failure can break the graph into two subgraphs. However, taking into account all other benefits of DPF methods (especially, scalability), they are still preferred over leader-agent methods in tree graphs. 

\section{Conclusion}\label{sec:conc}
We have studied DPF for target tracking and compared five consensus methods in a unified scenario, in terms of RMSE performance and communication cost. The five methods include four from literature (SBC, RG, MBC, and BG), and one novel method based on BP. According to our results, DPF-MBC should be used in loopy networks, while DPF-BP is preferred in tree networks. We also found that BC-based DPF methods have lower communication overhead than NCPF (based on data flooding) in sparse networks. Further research is required to estimate the diameter of the graph in distributed way, to reduce the bias of DPF-BP in loopy networks, and to assess the impact of medium access control on the communication delay.

\appendix  
\section{Derivation of BP Consensus}\label{subsec:bp-eq}
From (\ref{msgs2short}), we know that
\be
\label{msgs2un}
m_{un}^{(i)} (\mathbf{x}_t) \propto \frac{{M_u^{(i - 1)} (\mathbf{x}_t)}}{{m_{nu}^{(i - 1)} (\mathbf{x}_t)}},
\ee
where we removed index \emph{n} since all the nodes have the same variable ($\mathbf{x}_{n,t}=\mathbf{x}_{u,t}=\mathbf{x}_t$). The denominator of (\ref{msgs2un}) is the message from node $n$ to node $u$ in the previous iteration, and can be expressed as
\be
\label{msgs2nu}
m_{nu}^{(i-1)} (\mathbf{x}_t) \propto  \frac{{M_n^{(i - 2)} (\mathbf{x}_t)}}{{m_{un}^{(i - 2)} (\mathbf{x}_t)}}.
\ee
Combining previous two equations, we get the recursive expression for the messages
\be
\label{msgs3nu}
m_{un}^{{(i)}} (\mathbf{x}_t) \propto  \frac{{M_u^{(i-1)} (\mathbf{x}_t)}}{{M_n^{(i-2)} (\mathbf{x}_t)}}m_{un}^{(i-2)} (\mathbf{x}_t)
\ee

Combining (\ref{beliefs1}) and (\ref{msgs3nu}), we find a recursive expression for the beliefs: 
\begin{eqnarray}
\label{beliefs2}
M_n^{(i)} (\mathbf{x}_t) & \propto & p(y_{n,t}|\mathbf{x}_t)\prod_{u \in G_n } {\left(\frac{{M_u^{(i - 1)} (\mathbf{x}_t)}}{{M_n^{(i - 2)} (\mathbf{x}_t)}}m_{un}^{(i - 2)} (\mathbf{x}_t)\right)} \nonumber \\
& = & M_n^{(i-2)} (\mathbf{x}_t) \prod\limits_{u \in G_n }{\left(\frac{{M_u^{(i - 1)} (\mathbf{x}_t)}}{M_n^{(i - 2)} (\mathbf{x}_t)}\right)}.
\end{eqnarray}

\section{Convergence Behavior of BP Consensus in Loopy Graphs}\label{subsec:bp-corr}

In this appendix, we provide a deeper analysis of convergence behavior of BP consensus in loopy graphs. It is  well-known \cite{Pearl1988} that BP consensus (as a special case of standard BP) 
converges to the exact solution after a finite number of iterations in cycle-free graphs. Using an appropriate message schedule, this number of iterations is equal to $D_g+1$, where $D_g$ is the diameter of the graph (i.e., the maximum hop-distance between any two nodes). However, for general graphs, it is straightforward to show (using equation (\ref{new-bc})) that the beliefs of BP consensus after $D_g+1$ iterations is given by
\be
\label{eq:bp-estimate}
M_n^{(D_g+1)}(\mathbf{x}_t) \propto \prod_{u \in G_t }p({y_{u,t}}|{\mathbf{x}_{t}})^{\alpha_{u,n,t}}
\ee
where $\alpha_{u,n,t}\geq 1$ is an exponent ($\alpha_{u,n,t} \in \mathbb{N}$) of node pair $(u,n)$ at time \emph{t}. In case of cycle-free graphs and some specific symmetric graphs (see later examples), $\alpha_{u,n,t}= 1$, so the estimated belief is equal to desired global likelihood (given by (\ref{eq:g-lhood})). That means that DPF-BP, after  $D_g+1$ iterations, provides (at each node) an estimate exactly the same as CPF/NCPF. In case of $\alpha_{u,t,n}> 1$, the observation from node \emph{u} at time \emph{t} is \emph{over-counted} at node $n$. To understand the overcounting behavior, we determine $\alpha_{\mathrm{max}}$, the maximum value (maximized over $n$ and $u$) of $\alpha_{u,n,t}$  after $D_g+1$ iterations.  Note that running more than $D_g+1$ iterations is unnecessary, as it will increase the $\alpha$-values. While for the general case this problem is hard, we limit ourselves to some  best- and the worst-case examples. In particular, we consider 4 representative graph configurations, shown in Figure \ref{fig:graphs-bp}:

\begin{figure}[th]
\centering
\includegraphics[width=0.4\textwidth]{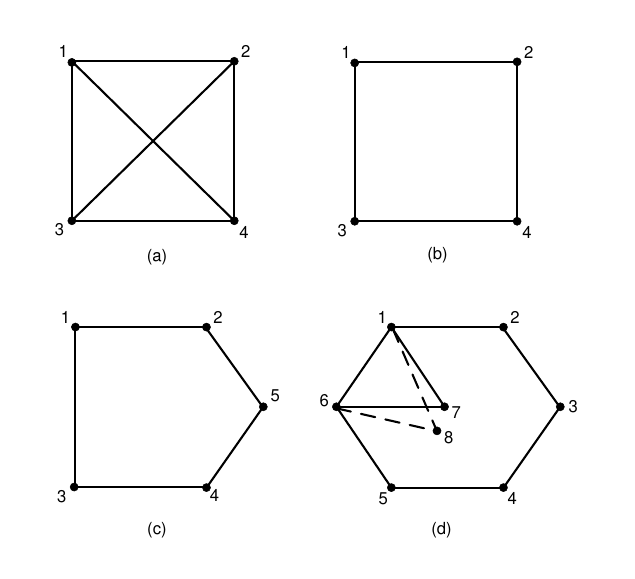} 
\caption{Example graphs: (a) fully-connected graph ($D_g=1$), (b) single-cycle graph with even number of nodes ($D_g=2$), (c) single-cycle graph with odd number of nodes ($D_g=2$), and (d) single-cycle graph with added short loop(s) ($D_g=3$). }
\label{fig:graphs-bp}
\end{figure}

\bn
\item \textit{Fully-connected graph (clique):} For the example in Figure \ref{fig:graphs-bp}a, $D_g=1$, so the belief at second iterations is given by (\ref{semi-local-init}). Since the graph is fully-connected, we know that set $G_n$ includes all nodes in the graph except node \emph{n} (which is locally available). Therefore, $\alpha_{\rm max}=1$, so BP consensus is correct. 

\item \textit{Single-cycle graph with even number of nodes:} For the example in Figure \ref{fig:graphs-bp}b, $D_g=2$, so we need to run 3 iterations of BP. In the second iteration, node 1 will obtain likelihood from nodes 2 and 3, but in the third iteration it will obtain likelihood from node 4 twice (through nodes 2 and 3). Therefore, $\alpha_{\rm max}=2$.

\item \textit{Single-cycle graph with odd number of nodes:} For the example in Figure \ref{fig:graphs-bp}c, again $D_g=2$, so we need to run 3 iterations of BP. In the second iteration, node 1 will obtain likelihood from nodes 2 and 3, and in the third iteration it will obtain likelihood from nodes 4 and 5. Therefore, $\alpha_{\rm max}=1$,  so BP consensus is correct. 

\item \textit{Graph with short loops:} For the example in Figure \ref{fig:graphs-bp}d (without node 8), $D_g=3$, so we need 4 iterations of BP. After 4 iterations, nodes 1 and 6 will have triple-counted their own local likelihoods (since it has its own information, as well as messages received due to the clockwise and counter-clockwise circulation through short loop\footnote{A short loop is defined as a loop that consists of 3 nodes.} 1-6-7). Therefore, $\alpha_{\rm max}=3$. If we consider the case with node 8 and two dashed links (in Figure \ref{fig:graphs-bp}d), $\alpha_{\rm max}=5$. This reasoning can be generalized to a case with $N_{\rm short-loops}$ short loops (which all contain the link 1-6), $\alpha_{\rm max}=1+2N_{\rm short-loops}$.

\en

\begin{table}[tb]
\renewcommand{\arraystretch}{1.3}
\caption{Estimates of BP consensus (first 3 iterations) for the graph in Figure \ref{fig:graphs-bp}c. The local likelihood $p({y_{u,t}}|{x_{t}})$ is marked by $\phi_{u,t}$.}
\label{table:bp-cycle-odd}
\centering
\begin{tabular}{c c  c  c}
 & iter. 1 & iter. 2 & iter. 3\\
\hline\hline
node 1 & $\phi_{1,t}$ & $\phi_{1,t}\phi_{2,t}\phi_{3,t}$ & $\phi_{1,t}\phi_{2,t}\phi_{3,t}\phi_{4,t}\phi_{5,t}$\\ \hline
node 2 & $\phi_{2,t}$ & $\phi_{1,t}\phi_{2,t}\phi_{5,t}$ & $\phi_{1,t}\phi_{2,t}\phi_{3,t}\phi_{4,t}\phi_{5,t}$\\ \hline
node 3 & $\phi_{3,t}$ & $\phi_{1,t}\phi_{3,t}\phi_{4,t}$ & $\phi_{1,t}\phi_{2,t}\phi_{3,t}\phi_{4,t}\phi_{5,t}$\\ \hline
node 4 & $\phi_{4,t}$ & $\phi_{3,t}\phi_{4,t}\phi_{5,t}$ & $\phi_{1,t}\phi_{2,t}\phi_{3,t}\phi_{4,t}\phi_{5,t}$\\ \hline
node 5 & $\phi_{5,t}$ & $\phi_{2,t}\phi_{4,t}\phi_{5,t}$ & $\phi_{1,t}\phi_{2,t}\phi_{3,t}\phi_{4,t}\phi_{5,t}$\\
\end{tabular}
\end{table}

All previous claims can be easily proved using (\ref{new-bc}) and (\ref{semi-local-init}). As example, we show in Table \ref{table:bp-cycle-odd} the estimates for the cycle-graph with odd number of nodes.

Taking into account that the fourth case is the worst-case scenario, we can conclude that in the worst-case $\alpha_{\rm max}=1+2N_{\rm short-loops}$. This is not a promising conclusion, since $\alpha_{\rm max}$ can be unbounded, for fixed $D_g$, as the number of nodes grows. Therefore, the configurations in which there are many short loops over the same link, are not preferable for BP consensus.



\begin{thebibliography}{10}

\bibitem{Hlinka2013}
O.~Hlinka, F.~Hlawatsch, and P.~M. Djuric, ``Distributed particle filtering in
  agent networks: A survey, classification, and comparison,'' {\em IEEE Signal
  Processing Magazine}, vol.~30, pp.~61--81, Jan. 2013.

\bibitem{Welch2006}
G.~Welch and G.~Bishop, ``An introduction to the {K}alman filter,'' tech. rep.,
  University of North Carolina at Chapel Hill, July 2006.

\bibitem{Arulampalam2002}
M.~S. Arulampalam, S.~Maskell, N.~G. Gordon, and T.~Clapp, ``A tutorial on
  particle filters for online nonlinear/non-{G}aussian {B}ayesian tracking,''
  {\em IEEE Transactions on Signal Processing}, vol.~50, pp.~174--188, Feb.
  2002.

\bibitem{Coates2004}
M.~Coates, ``Distributed particle filters for sensor networks,'' in {\em Proc.
  of 3rd Workshop on Information Processing in Sensor Networks (IPSN)},
  pp.~99--107, April 2004.

\bibitem{Sheng2005}
X.~Sheng, Y.-H. Hu, and P.~Ramanathan, ``Distributed particle filter with {GMM}
  approximation for multiple targets localization and tracking in wireless
  sensor network,'' in {\em Proc. of Fourth Int. Symp. Information Processing
  in Sensor Networks (IPSN)}, pp.~181--188, April 2005.

\bibitem{Lee2009}
S.~H. Lee and M.~West, ``Markov chain distributed particle filters ({MCDPF}),''
  in {\em Proc. of 48th IEEE Conf. held jointly with the 2009 28th Chinese
  Control Conf Decision and Control (CDC/CCC)}, pp.~5496--5501, Dec. 2009.

\bibitem{Hlinka2009}
O.~Hlinka, P.~M. Djuric, and F.~Hlawatsch, ``Time-space-sequential distributed
  particle filtering with low-rate communications,'' in {\em Proc. of Asilomar
  Conf.}, pp.~196--200, Nov. 2009.

\bibitem{Coates2005}
M.~Coates and G.~Ing, ``Sensor network particle filters: motes as particles,''
  in {\em in Proc. of IEEE Workshop on Statistical Signal Processing (SSP)},
  July 2005.

\bibitem{Jiang2011}
B.~Jiang and B.~Ravindran, ``Completely distributed particle filters for target
  tracking in sensor networks,'' in {\em Proc. of IEEE Int. Parallel \&
  Distributed Processing Symp.}, pp.~334--344, May 2011.

\bibitem{Djuric2011a}
P.~M. Djuric, J.~Beaudeau, and M.~Bugallo, ``Non-centralized target tracking
  with mobile agents,'' in {\em Proc. of IEEE Int. Conf. on Acoustics, Speech
  and Signal Processing (ICASSP)}, pp.~5928--5931, May 2011.

\bibitem{Gu2007a}
D.~Gu, ``Distributed particle filter for target tracking,'' in {\em Proc. of
  IEEE Int. Conf. on Robotics and Automation (ICRA)}, pp.~3856--3861, April
  2007.

\bibitem{Gu2008}
D.~Gu, J.~Sun, Z.~Hu, and H.~Li, ``Consensus based distributed particle filter
  in sensor networks,'' in {\em Proc. of Int. Conf. Information and
  Automation}, pp.~302--307, June 2008.

\bibitem{Hlinka2011}
O.~Hlinka, O.~Sluciak, F.~Hlawatsch, P.~M. Djuric, and M.~Rupp, ``Distributed
  gaussian particle filtering using likelihood consensus,'' in {\em Proc. of
  IEEE Int. Conf. on Acoustics, Speech and Signal Processing (ICASSP)},
  pp.~3756--3759, May 2011.

\bibitem{Hlinka2012}
O.~Hlinka, O.~Sluciak, F.~Hlawatsch, P.~Djuric, and M.~Rupp, ``Likelihood
  consensus and its application to distributed particle filtering,'' {\em IEEE
  Transactions on Signal Processing}, vol.~60, pp.~4334--4349, Aug. 2012.

\bibitem{Oreshkin2010}
B.~N. Oreshkin and M.~J. Coates, ``Asynchronous distributed particle filter via
  decentralized evaluation of gaussian products,'' in {\em Proc. of 13th Conf.
  on Information Fusion (FUSION)}, pp.~1--8, July 2010.

\bibitem{Ustebay2011}
D.~Ustebay, M.~Coates, and M.~Rabbat, ``Distributed auxiliary particle filters
  using selective gossip,'' in {\em Proc. of IEEE Int. Conf. on Acoustics,
  Speech and Signal Processing (ICASSP)}, pp.~3296--3299, May 2011.

\bibitem{Farahmand2011}
S.~Farahmand, S.~I. Roumeliotis, and G.~B. Giannakis, ``Set-membership
  constrained particle filter: Distributed adaptation for sensor networks,''
  {\em IEEE Trans. on Signal Processing}, vol.~59, pp.~4122--4138, Sept.
  2011.
    
\bibitem{Lindberg2013}
C.~Lindberg, L.~S.~Muppirisetty, K-M.~Dahl\'en, V.~Savic, and H.~Wymeersch. ``MAC Delay in Belief Consensus for Distributed Tracking,'' in {\em Proc. of 10th Workshop on Positioning, Navigation and Communication (WPNC)}, March 2013. 

\bibitem{Liu2009}
H.~Q. Liu, H.~C. So, F.~K.~W. Chan, and K.~W.~K. Lui, ``Distributed particle
  filter for target tracking in sensor networks,'' {\em Progress In
  Electromagnetics Research}, vol.~11, pp.~171--182, 2009.

\bibitem{Savic2012eusip}
V.~Savic, H.~Wymeersch, and S.~Zazo, ``Distributed target tracking based on
  belief propagation consensus,'' in {\em Proc. of the 20th European Signal
  Processing Conference (EUSIPCO)}, pp.~544--548, Aug. 2012.

\bibitem{Savic2011a}
V.~Savic, A.~Athalye, M.~Bolic, and P.~M. Djuric, ``Particle filtering for
  indoor {RFID} tag tracking,'' in {\em Proc. of IEEE Statistical Signal
  Processing Workshop (SSP)}, pp.~193 --196, June 2011.

\bibitem{Ahmed2010}
N.~Ahmed, M.~Rutten, T.~Bessell, S.~S. Kanhere, N.~Gordon, and S.~Jha,
  ``Detection and tracking using particle-filter-based wireless sensor
  networks,'' {\em IEEE Transactions on Mobile Computing}, vol.~9,
  pp.~1332--1345, Sept. 2010.

\bibitem{Oka2010}
A.~Oka and L.~Lampe, ``Distributed target tracking using signal strength
  measurements by a wireless sensor network,'' {\em IEEE Journal on Selected
  Areas in Communications}, vol.~28, pp.~1006--1015, Sept. 2010.

\bibitem{Chen2011}
X.~Chen, A.~Edelstein, Y.~Li, M.~Coates, M.~Rabbat, and A.~Men, ``Sequential
  {M}onte {C}arlo for simultaneous passive device-free tracking and sensor
  localization using received signal strength measurements,'' in {\em Proc. of
  IEEE/ACM Int. Conf. on Information Processing in Sensor Networks (IPSN)},
  pp.~342--353, April 2011.

\bibitem{Ihler2005a}
A.~T. Ihler, J.~W.~I. Fisher, R.~L. Moses, and A.~S. Willsky, ``Nonparametric
  belief propagation for self-localization of sensor networks,'' {\em IEEE
  Journal on Selected Areas in Communications}, vol.~23, pp.~809--819, April
  2005.

\bibitem{Galstyan2004}
A.~Galstyan, B.~Krishnamachari, K.~Lerman, and S.~Pattem, ``Distributed online
  localization in sensor networks using a moving target,'' in {\em Proc. of 3rd
  Int. Symp. on Information Processing in Sensor Networks (IPSN)}, pp.~61--70,
  April 2004.

\bibitem{Pitt1999}
M.~K. Pitt and N.~Shephard, ``Filtering via simulation: Auxiliary particle
  filters,'' {\em Journal of the American Statistical Association}, vol.~94,
  pp.~590--599, June 1999.

\bibitem{Merwe2001}
R.~van~der Merwe, A.~Doucet, N.~D. Freitas, and E.~Wan, ``The unscented
  particle filter,'' in {\em Proc. of Advances in Neural Information Processing
  Systems}, Nov. 2001.

\bibitem{Kotecha2003}
J.~H. Kotecha and P.~M. Djuric, ``Gaussian sum particle filtering,'' {\em IEEE
  Transactions on Signal Processing}, vol.~51, pp.~2602--1612, Oct. 2003.

\bibitem{Patwari2005}
N.~Patwari, J.~N. Ash, S.~Kyperountas, A.~O. Hero, III, R.~L. Moses, and N.~S.
  Correal, ``Locating the nodes: cooperative localization in wireless sensor
  networks,'' {\em IEEE Signal Processing Magazine}, vol.~22, pp.~54--69, July
  2005.

\bibitem{Olfati-Saber2004}
R.~Olfati-Saber and R.~Murray, ``Consensus problems in networks of agents with
  switching topology and time-delays,'' {\em IEEE Transactions on Automatic
  Control}, vol.~49, pp.~1520 -- 1533, Sept. 2004.

\bibitem{Leng2011}
M.~Leng and Y.-C. Wu, ``Distributed clock synchronization for wireless sensor
  networks using belief propagation,'' {\em IEEE Transactions on Signal
  Processing}, vol.~59, pp.~5404 --5414, Nov. 2011.

\bibitem{Olfati-saber2006}
R.~Olfati-Saber, E.~Franco, E.~Frazzoli, and J.~S. Shamma, ``Belief consensus
  and distributed hypothesis testing in sensor networks,'' in {\em Proc. of
  NESC Worskhop}, pp.~169--182, Springer Verlag, 2006.

\bibitem{Aysal2009}
T.~C. Aysal, M.~E. Yildiz, A.~D. Sarwate, and A.~Scaglione, ``Broadcast gossip
  algorithms for consensus,'' {\em IEEE Transactions on Signal Processing},
  vol.~57, pp.~2748--2761, July 2009.

\bibitem{Dimakis2010}
A.~G. Dimakis, S.~Kar, J.~M.~F. Moura, M.~G. Rabbat, and A.~Scaglione, ``Gossip
  algorithms for distributed signal processing,'' {\em Proc. of the IEEE},
  vol.~98, pp.~1847--1864, Nov. 2010.

\bibitem{Crick2003}
C.~Crick and A.~Pfeffer, ``Loopy belief propagation as a basis for
  communication in sensor networks,'' in {\em Uncertainty in Artificial
  Intelligence}, pp.~159--166, Aug. 2003.

\bibitem{Boyd2006}
S.~Boyd, A.~Ghosh, B.~Prabhakar, and D.~Shah, ``Randomized gossip algorithms,''
  {\em IEEE Transactions on Information Theory}, vol.~52, pp.~2508--2530, June
  2006.

\bibitem{Xiao2004}
L.~Xiao and S.~Boyd, ``Fast linear iterations for distributed averaging,'' {\em
  Systems and Control Letters}, vol.~53, no.~1, pp.~65 -- 78, 2004.

\bibitem{Pearl1988}
J.~Pearl, {\em Probabilistic Reasoning in Intelligent Systems: Networks of
  Plausible Inference}.
\newblock San Mateo: Morgan Kaufmann, 1988.

\bibitem{Xiao2007}
L.~Xiao, S.~Boyd, and S.-J. Kim, ``Distributed average consensus with
  least-mean-square deviation,'' {\em Journal of Parallel and Distributed
  Computing}, vol.~67, pp.~33--46, 2007.

\bibitem{Johansson2008}
B.~Johansson, {\em On Distributed Optimization in Networked Systems}.
\newblock PhD thesis, KTH, Stockholm, Sweden, Dec. 2008.

\bibitem{Aoyama2010}
D.~Mosk-Aoyama, T.~Roughgarden, and D.~Shah, ``Fully distributed algorithms for
  convex optimization problems,'' {\em SIAM Journal on Optimization}, vol.~20,
  pp.~3260--3279, Oct. 2010.

\bibitem{Pham2009}
T.-D. Pham, H.~Q. Ngo, V.-D. Le, S.~Lee, and Y.-K. Lee, ``Broadcast gossip
  based distributed hypothesis testing in wireless sensor networks,'' in {\em
  Proc. of Int. Conf. on Advanced Technologies for Communications}, pp.~84--87,
  Oct. 2009.

\bibitem{Yedidia2003}
J.~S. Yedidia, W.~T. Freeman, and Y.~Weiss, {\em Understanding belief
  propagation and its generalizations}, pp.~239--269.
\newblock San Francisco, CA, USA: Morgan Kaufmann Publishers Inc., 2003.

\bibitem{Weiss1998}
Y.~Weiss, ``Correctness of local probability propagation in graphical models
  with loops,'' {\em Neural Computation}, vol.~12, pp.~1--41, Jan. 2000.

\bibitem{Savic2013fusion}
V.~Savic, H.~Wymeersch and E.~G.~Larsson, ``Simultaneous sensor localization and target tracking in mine tunnels,'' {\em in IEEE Proc. of Intl. Conf. on Information Fusion}, July 2013.
  
\bibitem{Chehri2012}
A.~Chehri, P.~Fortier, and P.~M. Tardif, ``Characterization of the
  ultra-wideband channel in confined environments with diffracting rough
  surfaces,'' {\em Wireless Personal Communications (Springer)}, vol.~62,
  pp.~859--877, Feb. 2012.


\end{thebibliography}

\end{document}